\documentclass[final,english]{bullsrsl}[2022/06/15]

\usepackage[latin1]{inputenc}
\usepackage[T1]{fontenc}

\usepackage{natbib} % Add "numbers" option if natbib is used
\usepackage{graphicx}

\newcommand\aaps{\ref@jnl{A\&AS}}
\newcommand\actaa{\ref@jnl{Acta Astron.}}

\begin{document}
\title{Asteroseismology of hot subdwarf B stars observed with TESS: discovery of two new gravity mode pulsating stars}

\author[affil={1,2}, corresponding]{Murat}{Uzundag}
\author[affil={3}]{Roberto}{Silvotti}
\author[affil={4,5}]{Andrzej S.}{Baran}
\author[affil={1}]{Maja}{Vu\v{c}kovi\'{c}}
\author[affil={6}]{P\'eter}{N\'emeth}
\author[affil={4,7}]{Sumanta K.}{Sahoo}
\author[affil={8}]{Mike}{Reed}

\affiliation[1]{Instituto de F\'isica y Astronom\'ia, Universidad de Valpara\'iso, Gran Breta\~na 1111, Playa Ancha, Valpara\'iso 2360102, Chile}
\affiliation[2]{European Southern Observatory, Alonso de Cordova 3107, Santiago, Chile}
\affiliation[3]{INAF-Osservatorio Astrofisico di Torino, strada dell'Osservatorio 20, 10025 Pino Torinese, Italy}
\affiliation[4]{ARDASTELLA Research Group}
\affiliation[5]{Astronomical Observatory, University of Warsaw, Al. Ujazdowskie 4, 00-478 Warszawa, Poland}
\affiliation[6]{Astroserver.org, F\H{o} t\'er 1, 8533 Malomsok, Hungary}
\affiliation[7]{Nicolaus Copernicus Astronomical Centre of the Polish Academy of Sciences, ul. Bartycka 18, 00-716 Warsaw, Poland}
\affiliation[8]{Department of Physics, Astronomy and Materials Science, Missouri State University, 901 S. National, Springfield, MO 65897, USA}

\correspondance{muratuzundag.astro@gmail.com}
\date{13th October 2020}
\maketitle

% \author[affil1]{FirstName (+ MiddleInitials if necessary)}{FamilyName}
% \author[affil2]{...}{}
% \equalcontribauthor[]{}{} % Maximum two --> counter
% \consortium[affil]{Consortium Name}
% With consortium: affiliation will be set to "See Appendix 1 for a full
% list of consortium members and their respective affiliations
% \affiliation[affil1]{...}
% \affiliationq[affil2]{...}

% \correspondence[]{}
% No explicit corresponding author: use first author
% 

% Abstract of the paper in the same language as the paper
\begin{abstract}

   TIC\,033834484 and TIC\,309658435 are long-period pulsating subdwarf B star, which were observed extensively (675 and 621 days, respectively) by the Transiting Exoplanet Survey Satellite (TESS).
   The high-precision photometric light curve reveals the presence of more than 40 pulsation modes including both stars.
   All the oscillation frequencies that we found are associated with gravity (g)-mode pulsations, with frequencies spanning from $\sim$80 $\mu$Hz (2\,500 s) to $\sim$400 $\mu$Hz (12\,000 s).
   We utilize the asteroseismic tools including asymptotic period spacings and rotational frequency multiplets in order to identify the pulsational modes.
   We found dipole ($l = 1$) mode sequences for both targets and calculate the mean period spacing of dipole modes ($\Delta P_{l=1}$), which allows us to identify the modes. 
   Frequency multiplets provide a rotation period of $\sim$64 d for TIC\,033834484. 
   From follow-up ground-based spectroscopy, we find that TIC\,033834484 has an effective temperature of 24\,210\,K (140), a surface gravity of $\log\,g /cm s^{-2}$ = 5.28 (03) and TIC\,309658435 has an effective temperature of 25\,910\,K (150), a surface gravity of $\log\,g /cm s^{-2}$ = 5.48 (03).
   
\end{abstract}

\keywords{asteroseismology --- stars: oscillations (including pulsations) --- stars: interiors --- stars:  evolution --- stars: horizontal-branch  --- stars: subdwarfs --- Stars: individual:  TIC\,033834484, TIC\,309658435}

\section{Introduction}

%Hot subdwarf stars are core helium-burning stars with a very thin hydrogen envelope ($M_{env}$ $<$ 0.01 $M_{\odot}$), and a mass close to the core-helium-flash mass $\sim$0.47 \(\textup{M}_\odot\). The sdB stars are evolved compact (log $g$ = 5.0\,dex - 6.2\,dex) and hot (T$_{\rm eff}$ = 20\,000 - 40\,000\,K) objects with radii between 0.15 \(\textup{R}_\odot\) and 0.35 \(\textup{R}_\odot\), located on the so-called extreme horizontal branch \citep[EHB; see][for a review]{heber2016}.

One of the major progresses in our understanding of sdB stars was initiated by \citet{kilkenny1997} discovering rapid pulsations in hot sdBs known as V361 Hya stars (often referred to as short-period sdBV stars). The V361 Hya stars show multiperiodic pulsations with periods spanning from 60\,s to 800\,s. 
\citet{green2003} discovered the long-period sdB pulsators known as V1093 Her stars.
These stars show brightness variations with periods up to a few hours and have amplitudes smaller than 0.1 per cent of their mean brightness.
In the first group, the pulsational modes correspond to low-degree, low-order pressure (p)-modes while in the second group, the pulsational modes correspond to low-degree, medium- to high-order gravity (g)-modes.
Some sdB pulsators showing both g- and p-modes have been discovered among the two described families of pulsating sdB stars. These stars are referred to as hybrid sdB pulsating stars.
These modes are excited by a classical $\kappa$-mechanism due to the accumulation of the iron group elements (mostly iron itself), in the $Z$-bump region \citep{charpinet1996,charpinet1997,fontaine2003}. The authors also showed that radiative levitation is a key physical process to enhance the abundances of iron group elements in order to be able to excite the pulsational modes.  

With the advance of high precision ($\sim$ 0.02 $\mu$Hz) and high duty cycle ($>$ 90 \%) photometric monitoring from space, unprecedented asteroseismic measurements and tools have become available for sdB pulsators.
The non-radial oscillations observed in pulsating sdBs offer a unique way to probe these stars resolving their pulsation geometry, which is described by three quantized numbers: $l$ (modal degree) defines total surface nodes, $n$ (radial order) characterizes radial nodes from core to the surface and $m$ (azimuthal order) describes surface nodes which pass through the pulsation axis. 

For almost all the sdBVs observed from space, the asymptotic period sequences for g- mode pulsations have been successfully applied, especially for dipole and quadrupole modes, as more than 60\% of the periodicities are associated with these modes \citep{reed2011}. 
The asymptotic approximation can be perfectly applied for homogeneous stars. However, sdB stars are stratified and diffusion processes (gravitational settling and radiative levitation) contribute significantly to compositional discontinuities, which disturb the pulsational modes and could break the sequences. This effect has been shown in several sdBV stars. Furthermore, when the compositional discontinuities become stronger at the transition zones, some modes are trapped, and this effect was also seen in a few sdBV stars observed with Kepler.

Another asteroseismic tool, rotational multiplets, is useful to identify the pulsation modes as well as to determine the rotation period of the core and the surface of sdBVs. 
Single sdB stars tend to have a much longer rotation period from 16~d to 289~d (review by \citealt{Charpinet2018}, see also \citealt{Silvotti2022}). While sdBs in binaries have shorter rotation period between 2.42~h and 14.16~d.  

We are still discovering new sdBVs thanks to NASA's latest Transiting Exoplanet Survey Satellite (TESS), which is dedicated to high-precision photometric monitoring of stars from space.
Thus far, TESS has monitored thousand of sdB stars with 2-min and 20-sec cadence during 56 sectors. The stars at high latitude region, close to the
ecliptic caps, have been observed continuously (CVZ) and several sdBs were found in this zone. 
In this work, we present the analysis of two new discoveries of  hot subdwarf B pulsating stars. TIC\,033834484 and TIC\,309658435, were observed by TESS during the first and third year in the southern ecliptic hemisphere. 
We obtained a low-resolution spectra of each target and applied the model atmosphere models in order to derive the fundamental parameters of the stars. We performed the frequency analysis and detailed seismic mode identification for both targets.

\section{Spectroscopy}

Intermediate-resolution spectra (R = 3500) of TIC\,033834484 and TIC\,309658435 were collected during three half-nights of observation between 2015 January 28 and February 22 at Las Campanas Observatory, with the IMACS spectrograph at the
focus of the Baade 6.5 m telescope.
The instrument was used at f/4
in longslit mode, and the 0".75-wide slit was employed. 
The 1200$-$17.5 grating was tilted by an angle of $16^{o}.8$ to cover the spectral range 3660-5250 \AA\ on the CCD. 
Figure 1 shows the IMACS spectra together with their best-fit {\sc Tlusty/XTgrid} models.
%Both stars belong to the He-weak sdB spectroscopic group, which overlaps with g-mode sdBV pulsators and form a compact group on the EHB.
The determined spectroscopic parameters demonstrate that TIC\,033834484 has $T_{\rm eff} = 24210 (\pm140)$ (K) and $\log{g} = 5.28 (\pm0.03)$ (cm\,s$^{-2}$) while TIC\,309658435 has $T_{\rm eff} = 25910 (\pm150)$ (K) and $\log{g} = 5.47 (\pm0.03)$ (cm\,s$^{-2}$).
The spectra of both sdB stars are dominated by Balmer-lines and only weak He I lines are seen. 
Both stars belong to the He-weak sdB spectroscopic group, which overlaps with g-mode sdBV pulsators. The atmospheric parameters are in good agreement with the earlier results by \citet{MoniBidin2017}.

\begin{figure}
\centering
\includegraphics[width=13cm,height=10cm,keepaspectratio]{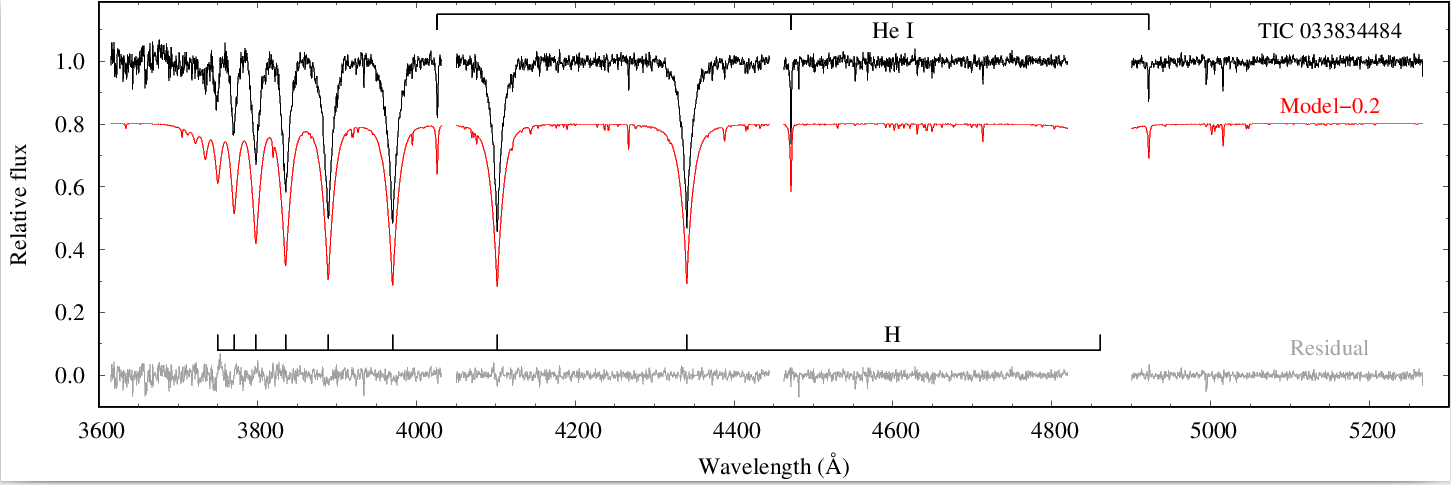}
\includegraphics[width=13cm,height=10cm,keepaspectratio]{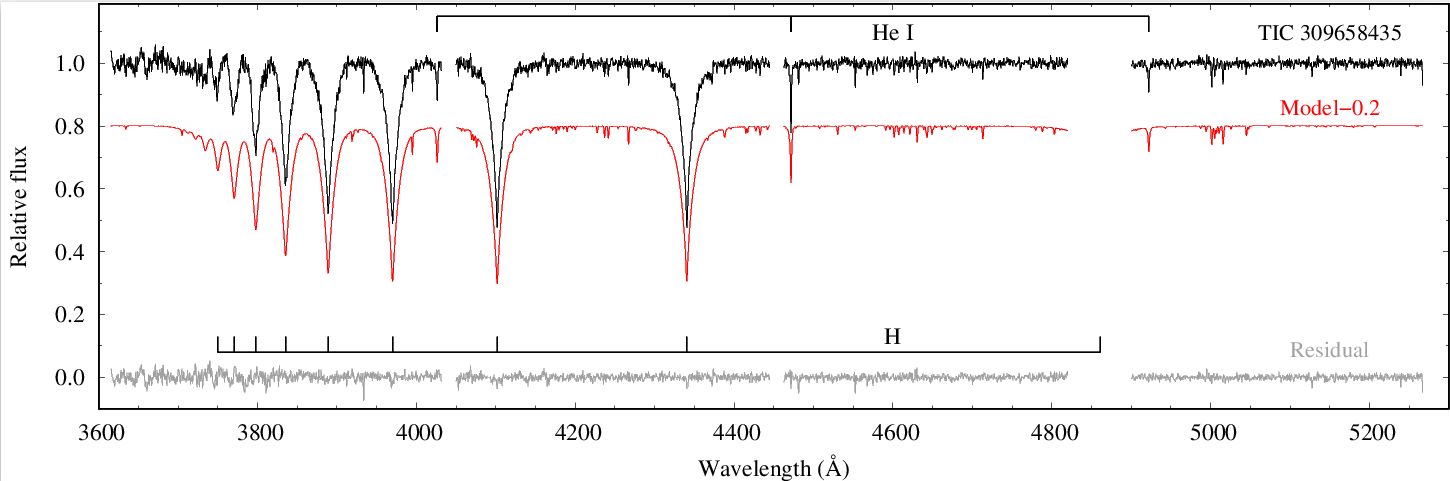}
\bigskip
\begin{minipage}{12cm}
\caption{Optical spectra of the two new pulsating hot subdwarf B pulsating stars (black lines). Overplotted are the best-fit atmospheric models (red lines). Identifications of Balmer-lines and weak He\,{\sc I} lines are marked.}
\label{fig:spectral_fits}
\end{minipage}
\end{figure}

\section{Photometric data}

%The TESS spacecraft observed the new pulsating hot subdwarf stars  and  (). {\bf IT IS WORTH MENTIONING THE STARS BEING IN THE CVZ}
TIC\,033834484 and TIC\,309658435 located in the TESS' CVZ, were observed extensively during 675 and 621 days, respectively. 
We analyzed  TESS observations using the short-cadence (SC) mode with 2-minutes sampling, allowing us to analyze the frequency range up to the Nyquist frequency at about 4167 $\mu$Hz. 
The Fourier transforms (FT) of the light curves were computed to examine the periodicities present in the data, aiming at identifying the frequency of all pulsation modes, along with their phase and amplitude.

\section{Asteroseismology}

\subsection{Rotational multiplets}

In rotating stars, the pulsation frequencies are split into 2$l +$1 azimuthal components due to rotation, revealing equally spaced multiplets of 2$l +$1 components. This 2$l +$1 configuration can be resolved with high-precision photometry if the star has no strong magnetic field and the rotational period is not longer than the duration of the observation.
Also, in order to detect rotationally split modes, the rotational axis has to be aligned
%[{\bf WHY? you can look at my cited paper (Silvotti+2022) and references therein for an answer to this question}] 
with the pulsation axis, otherwise the pulsation frequencies are split into nine components for $l = 1$ mode \citep[][and references therein]{Silvotti2022}.
Lastly, the inclination angle has to be different from 0 ($i \neq 0$). 

The following equation gives the rotation period (P$_{\rm rot}$) in terms of frequency splitting $\Delta \nu_{n,l,m}$.

\begin{equation}
\nu_{n,l,m} = \nu_{n,l,0} + \Delta \nu_{n,l,m} = \nu_{n,l,0} + m \frac{1- C_{n,l}}{P_{\rm rot}},
\end{equation}

where $C_{n,l}$ is the Ledoux constant \citep{Ledoux1951}, which, in the asymptotic limit, depends on the modal degree as $C_{n,l} = 1/l(l+1)$. For dipole and quadrupole modes therefore we get $C_{n,1} \sim$ 0.5 and $C_{n,2} \sim$ 0.17, respectively. 

\begin{figure}
\centering
\includegraphics[width=8cm,height=8cm,keepaspectratio]{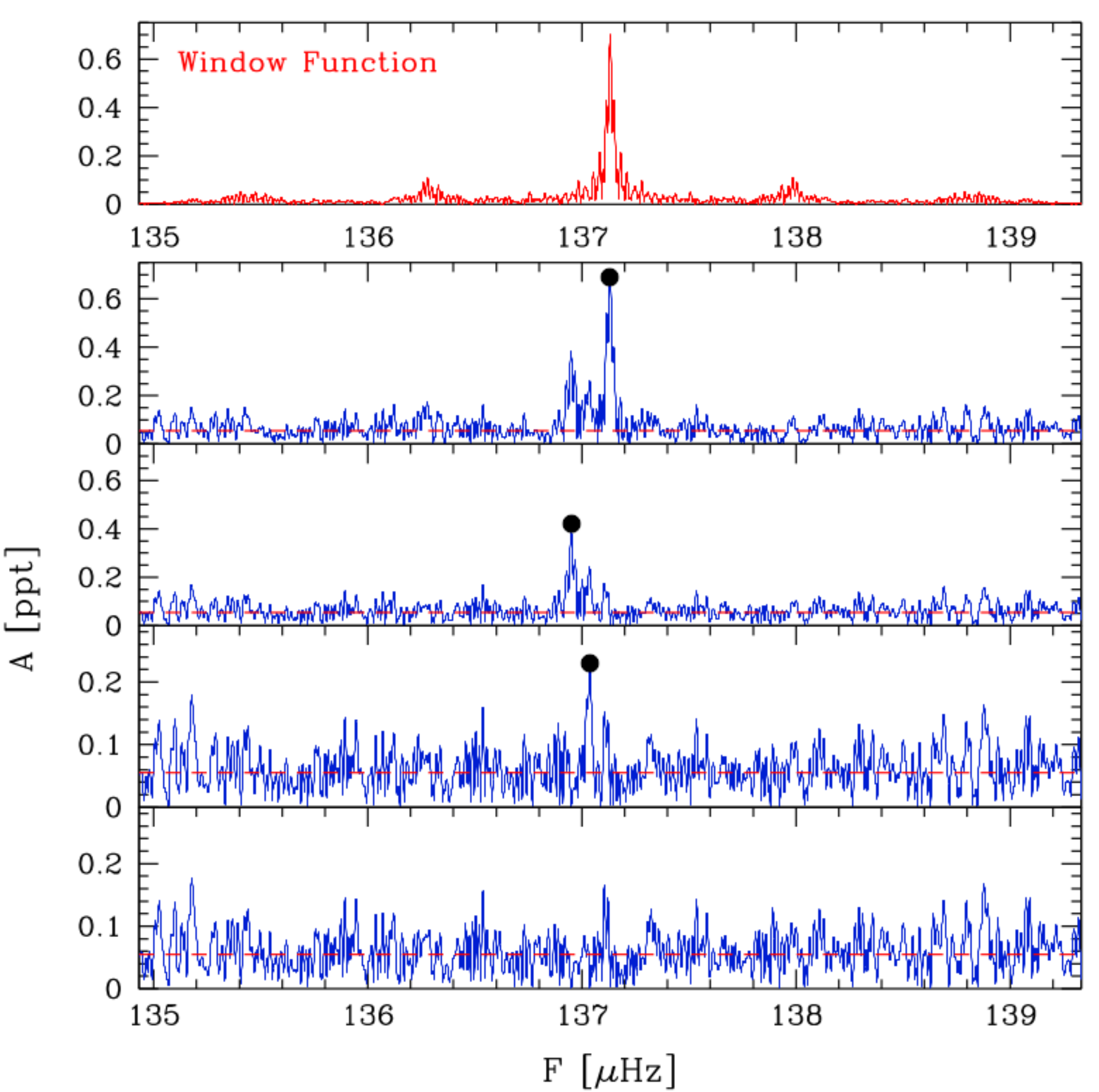}\includegraphics[width=8cm,height=8cm,keepaspectratio]{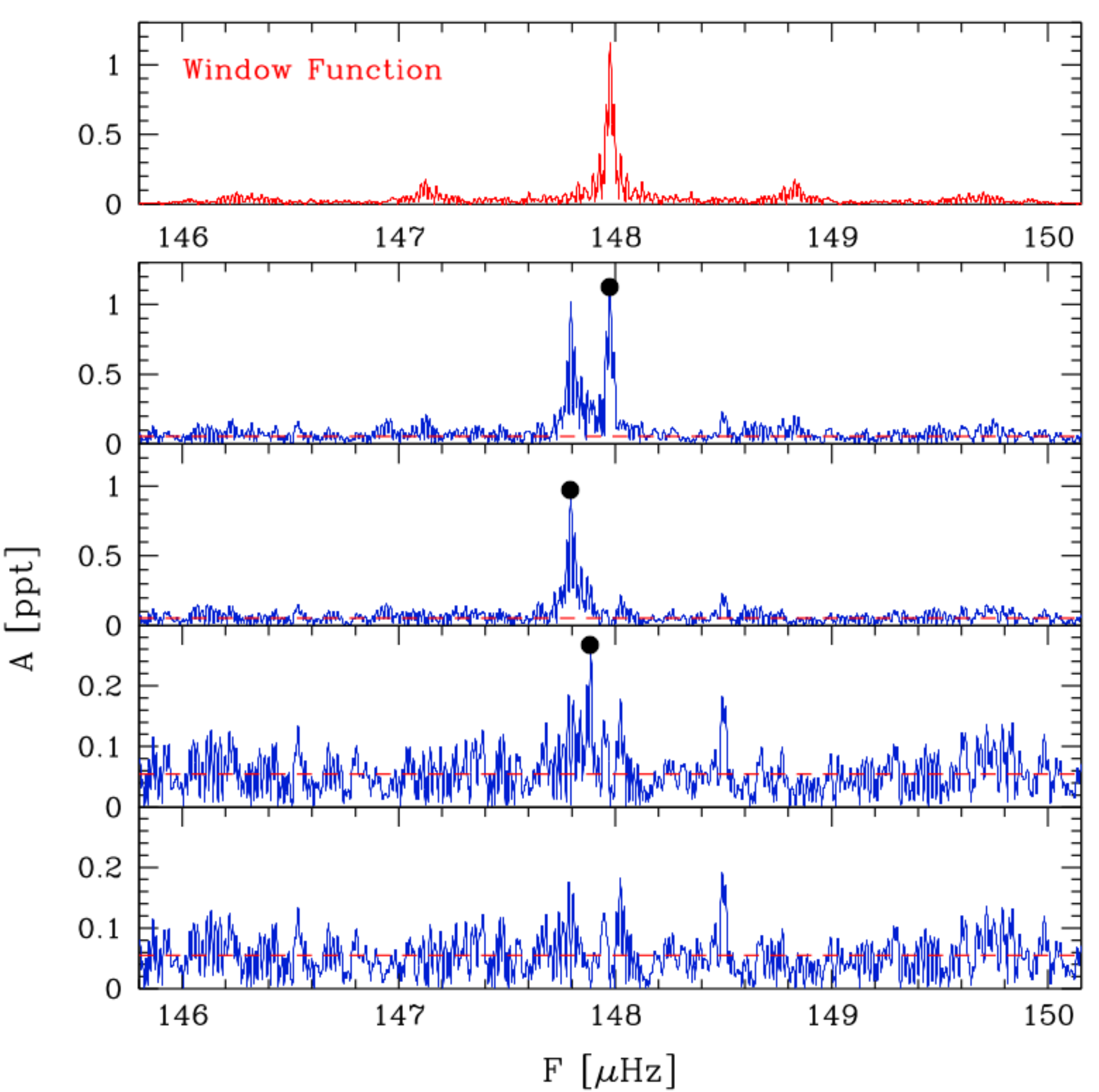}
\smallskip
\begin{minipage}{12cm}
\caption{FT of TIC\,033834484 showing the region around the frequency triplets near 137 $\mu$Hz (left panel) and 148 $\mu$Hz (right panel). From top to bottom we show
the window function, the original FT, and the FT of the residuals after subtracting each sequence of frequencies.
A black dot represents each prewhitened peak.}
\label{fig:ft2}
\end{minipage}
\end{figure}

For TIC\,033834484, we were able to determine a frequency splitting of $\sim$0.0905 $\mu$Hz
(Fig.\,\ref{fig:ft2}), and therefore this star appears to be another slow-rotating single sdB pulsating star with a rotation period of $\sim$64 days.

\subsection{Asymptotic period spacing}

\begin{figure}
\centering
\includegraphics[width=17cm,height=8cm,keepaspectratio]{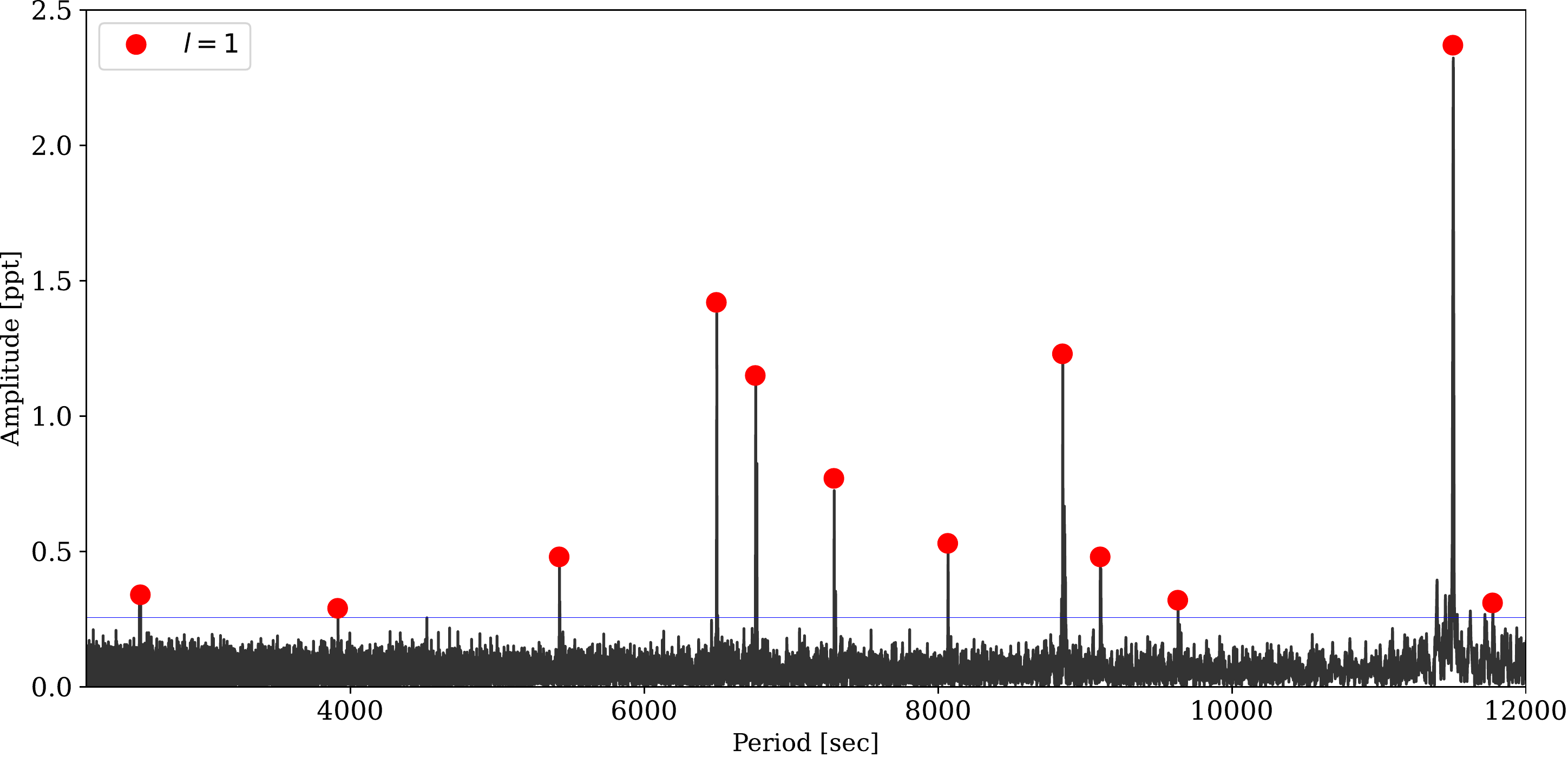}
\includegraphics[width=17cm,height=8cm,keepaspectratio]{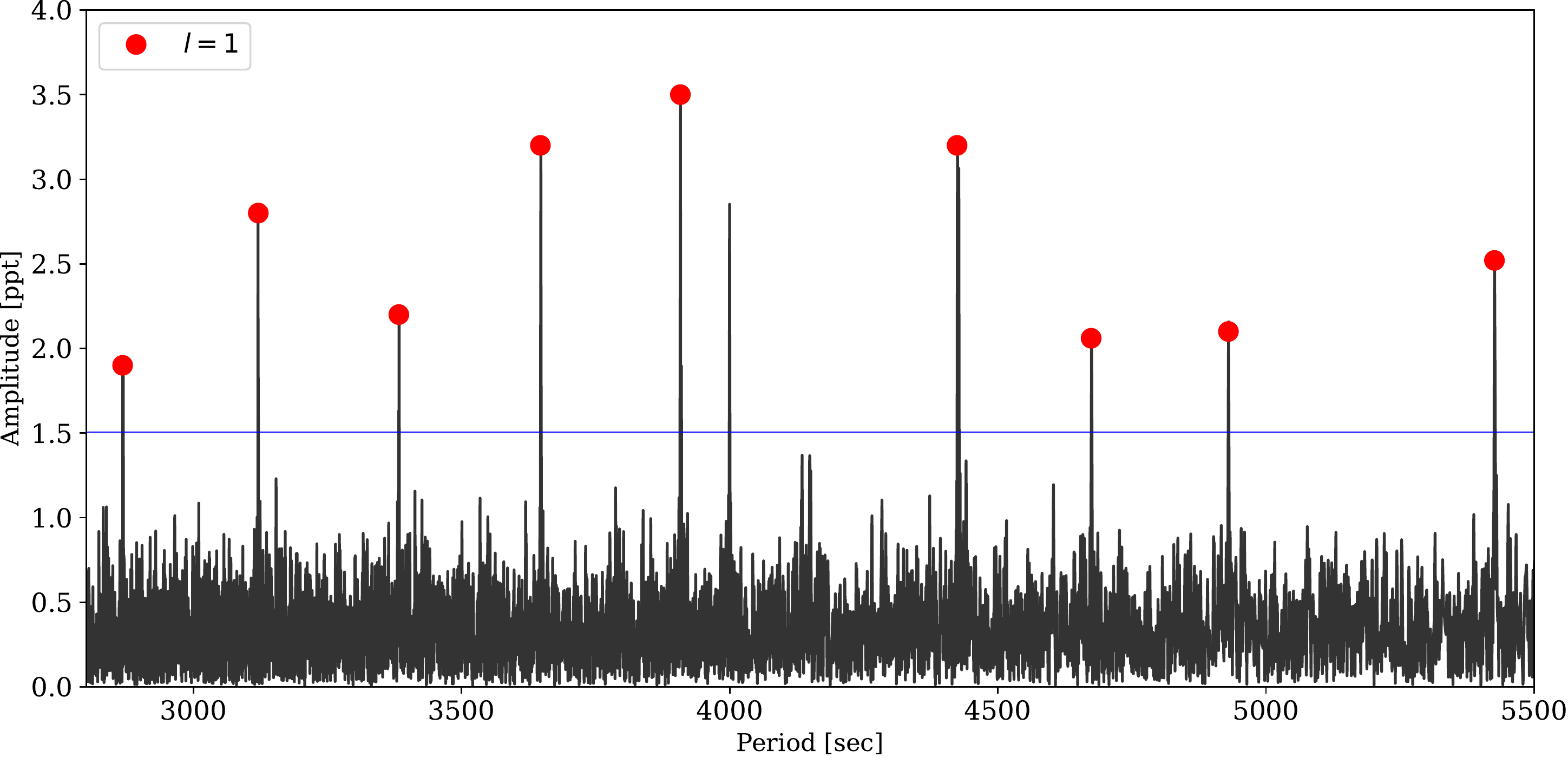}
\bigskip
\begin{minipage}{12cm}
\caption{Pulsation spectrum of TIC\,033834484 (top panel) and TIC\,309658435 (bottom panel) in the period space with the red dots indicating the expected locations of $\ell=1$ modes from the asymptotic pulsation theory. The horizontal blue lines show the confidence level of 0.1\% FAP.}
\label{fig:ft}
\end{minipage}
\end{figure}

In the asymptotic limit of stellar pulsations, i.e., for large radial orders 
($k \gg \ell$), $g$-modes of consecutive radial 
order in sdBs are approximately uniformly
spaced in period \citep{Tassoul1980}. The asymptotic period spacing
is given by $P_{\ell, n} = \Delta P_{0}/{\sqrt{\ell(\ell+1)}} n + \epsilon_{l}$,
where $\Delta P_0$ is the asymptotic period spacing for g-modes, which is defined as $\Delta P_{0} \propto  [\int_{r_1}^{r_2} \frac{|N|}{r} dr]^{-1}$, $N$ being the Brunt-V\"ais\"al\"a frequency, the critical frequency of nonradial g-modes \citep{Tassoul1980}, and
$\epsilon_{l}$ is a constant \citep{unno1989}.
The pulsation spectrum is displayed in units of period in Fig.\,\ref{fig:ft} with the expected locations of the $\ell=1$ modes for even period spacing indicated.
We derived the mean period spacing of dipole modes ($\Delta P_{l=1}$) and found $263.84^{+0.8}_{-1.4}$ s for TIC\,033834484, $257.06^{+1.5}_{-1.4}$ s for TIC\,309658435.
We used a bootstrap resampling analysis as described by \citet{uzundag2021} in order to propagate the errors in the mean period spacing obtained in our analysis.

\section{Results and conclusions}

We presented the discovery of two new
long-period sdB pulsating stars TIC\,033834484 and TIC\,309658435. The long dataset allows us to achieve a high frequency resolution of $\sim$0.02 $\mu$Hz and study the temporal stability of the pulsations, as well as to reveal the rotation period of TIC\,033834484.
We derived the atmospheric parameters for both stars by fitting synthetic models to the spectra.
We investigated the potential variability of the two new
sdBVs by examining their short and ultra-short-cadence observations obtained with TESS.
Our investigation of the detected frequencies reveals:

\begin{itemize}

\item{12 dipole ($l = 1$) modes for TIC\,033834484 with the relative radial order between 20 and 44 and mean period spacing of dipole modes ($\Delta P_{l=1}$) = $263.84^{+0.8}_{-1.4}$ s.}

\item{9 dipole modes for TIC\,309658435 with the relative radial order between 11 and 21 and mean period spacing of dipole modes ($\Delta P_{l=1}$) = $257.06^{+1.5}_{-1.4}$ s.}

\item{Rotational triplets for TIC\,033834484 with $\Delta \nu_{l = 1}$ of $\sim$0.0905 $\mu$Hz, corresponding to a stellar rotation period of about 64 days, making thi star another slowly rotating single pulsating sdB among ten others \citep{Silvotti2022}.
TIC\,309658435 does not show any multiplets suggesting that it has either a very slow rotation period (longer than $\sim$600 days) or a pole-on orientation of the pulsation axis, or the side components are not driven.}
%{\bf OR THE SIDE COMPONENTS ARE NOT DRIVEN}

\end{itemize}

%\bibliographystyle{bullsrsl-en}
%\bibliography{extra}

\begin{acknowledgments}
M.U. acknowledges financial support from CONICYT Doctorado Nacional in the form of grant number No: 21190886 and ESO studentship program.
P.N. acknowledges support from the Grant Agency of the Czech Republic (GA\v{C}R 22-34467S). 
The Astronomical Institute in Ond\v{r}ejov is supported by the project RVO:67985815.
This research has used the Sandbox services of \mbox{\url{www.Astroserver.org}}.
The authors warmly thank Christian MoniBidin, who provided the spectroscopic observation of the two targets.
Financial support from the National Science Center in Poland under project No.\,UMO-2017/26/E/ST9/00703 is acknowledged.

\end{acknowledgments}

\begin{furtherinformation}

\begin{orcids}

\orcid{0000-0002-6603-994X}{Murat}{Uzundag}
\orcid{0000-0003-0963-0239}{P\'eter}{N\'emeth} \\

\end{orcids}

\begin{authorcontributions}

M.U. and R.S. and A.S.B. carried out the light curve analyses as well as frequency solutions.
P.N. performed the atmospheric analysis. 
The text was written by M.U.
All authors contributed to the discussion and interpretation of the results and commented on the written draft of the paper.

\end{authorcontributions}

\begin{conflictsofinterest}

%This section is \emph{mandatory}. Authors must declare any personal or professional circumstances that may be perceived as influencing the research reported in the paper. If there is no conflict of interest, please state that ``

The authors declare no conflict of interest.

\end{conflictsofinterest}

\end{furtherinformation}

\bibliographystyle{bullsrsl-en}
%\bibliographystyle{bullsrsl-numen}

% If you use BibTeX, please list the bibliography databases in the
% \bibliography{...} command. Please notice that you will have to provide us
% with a (sub-setted) single file database once your manuscript has been
% accepted for publication and goes into production. Sub-setting can be easily
% done with bibexport.sh or bibtool, both available from CTAN.
\bibliography{extra.bib}

\end{document}